\documentstyle[11pt,newpasp,twoside,epsf]{article}

\begin{document}

\title{Coupling between satellite dwarfs and the Milky Way warp}
\author{Jeremy Bailin}
\affil{Steward Observatory, 933 N Cherry Ave, Tucson AZ, 85721 USA}
\author{Matthias Steinmetz}
\affil{Astrophysikalisches Institut Potsdam, An der Sternwarte 16, D-14482
  Potsdam, Germany}
\affil{Steward Observatory, 933 N Cherry Ave, Tucson AZ, 85721 USA}

\begin{abstract}
The perturbations of satellite galaxies, in particular the Large
Magellanic Cloud (LMC), have been repeatedly proposed and discounted
as the cause of the Milky Way warp. While the LMC may excite a
wake in the Galactic dark matter halo that could provide sufficient
torque to excite a warp of the observed magnitude, its orbit may be
incompatible with the orientation of the warp's line of nodes.
The Sgr dSph galaxy has
an appropriately-oriented orbit, and due to its closer orbit may produce
a stronger tidal effect than the LMC.
Evidence that Sgr may be responsible for the warp comes from its
orbital angular momentum, which has the same magnitude as the
angular momentum of the warped component
of the Galactic disk and is anti-aligned with it. We have run
high resolution numerical
N-body simulations of Sgr-sized satellites around a Milky Way-sized
disk to test this idea. 
Preliminary results suggest that Sgr-sized
satellites can indeed excite warps with properties similar 
to that observed for the Milky Way .
\end{abstract}

\section{Introduction}
Warped galactic disks are common phenomena; Reshetnikov~\&~Combes~(1998)
find that half of all disk galaxies have observable warps. This has been
a long-standing theoretical problem because isolated warps are not
long-lived (e.g.~Hunter~\& Toomre~1969).  
Therefore, there must either
be a method of stabilizing warps so they are long-lived, or they must
be generated
frequently.
The existence of apparently isolated strongly warped
galaxies, such as NGC~5907,
has led many authors  to
look for a universal method of stabilizing warps 
(e.g. Sparke~\& Casertano~1988;
L\'{o}pez-Corredoira, Betancort-Rijo,~\& Beckman~2002). 
However, with deeper
photometry
some of those galaxies, such as NGC~5907, now appear to be less isolated than
previously thought (Shang~et~al.~1998). Therefore, it may be that there
is no universal mechanism that stabilizes warps, but rather each
galaxy reacts to the specific perturbations 
caused by the mass distribution in its immediate environment.

The Milky Way is a warped galaxy, with the warp seen in neutral hydrogen
(Diplas~\& Savage 1991), dust (Freudenreich~et~al.~1994), and stars
(Drimmel, Smart,~\& Lattanzi~2000).
The Magellanic Clouds, as the
most massive perturbers in the Galactic neighborhood, are an
obvious candidate for causing the warp. Although Hunter~\&
Toomre~(1969) demonstrated that their direct torque is insufficient
to cause the Milky Way warp, Weinberg~(1998) proposed 
that the wake generated
by the LMC in the Galactic dark matter halo could amplify its effects.
Tsuchiya~(2002) has performed simulations that suggest that if the Milky Way's
halo is very massive, the LMC could indeed excite a large enough warp
after 6~Gyr. However, Garc\`{i}a-Ruiz, Kuijken,~\& Dubinski~(2002) noted that
a satellite on a fixed orbit generates a warp whose line of nodes
is in the plane of the satellite's orbit, while the Milky Way's line
of nodes is orthogonal to the LMC orbital plane.
While they did not take precession or sinking 
 of the satellite into account, the low resolution
but fully self-consistent simulations of Huang~\& Carlberg~(1997)
also indicate that as satellites sink, they lose angular momentum to
both the disk and the halo, which causes the disk to tilt and warp toward
(away) from the plane of the satellite's orbit for satellites in
prograde (retrograde) orbits.

The Sgr dSph, whose orbital plane intersects the Galactic line of nodes
(Ibata~et~al.~1997), might therefore be a good candidate for causing
the Milky Way's warp (Lin~1996). Although much less massive than the LMC,
its smaller galactocentric radius could strongly amplify its
effect (Bailin~2003).
Indeed, Ibata~\& Razoumov~(1998) have performed simulations in which they
passed a satellite on Sgr's orbit through a gas disk embedded
in a static Milky Way potential, and
found that for a massive enough satellite, the gas
disk was noticeably perturbed and warped.

In this proceeding, we examine some dynamical evidence that Sgr may be
responsible for the warp, and present some preliminary N-body simulations
of satellite-disk interactions with the aim of modelling the Sgr-Milky
Way system.

\section{Angular Momentum in the Warp and Satellites}
A torque is a transfer of angular momentum. A torque between two
components of a system will couple their angular momenta.

The angular momentum of the warped component of the disk can be estimated
by using a tilted ring model, where the angular momentum in a ring at radius
$R$ of width $dR$, surface density $\Sigma(R)$ and circular velocity
$v_c(R)$ has angular momentum
$$ dL = 2 \pi R^2\, v_c(R)\, \Sigma(R) \, dR $$
directed toward the rotation axis. The component of this which is
due to the warp, i.e. the component not perpendicular to the Galactic
plane, is (2--$8)\times 10^{12}~M_{\odot}~\mathrm{kpc~km~s^{-1}}$,
for a range of reasonable Galactic mass models (Bailin~2003).

The orbital angular momenta of satellite galaxies can be calculated
when they have measured proper motions.  The angular momentum of Sgr is
(2--$8)\times 10^{12}~M_{\odot}~\mathrm{kpc~km~s^{-1}}$, anti-aligned
with the warp angular momentum, while the other satellites have
angular momenta that span 3~orders of magnitude and cover a wide range
of directions (Bailin~2003). The coincidence of the
angular momenta of Sgr and of the Milky Way warp suggests that they
are coupled, i.e.~that Sgr is responsible for the warp.

\section{N-body Simulations}

Previous simulations of the interaction between Sgr and the
Milky Way disk have suffered from too low resolution
(Huang~\& Carlberg~(1997) had 80,000 particles in their galaxies while
 Weinberg~(1998) estimates that at least $10^6$ particles are necessary
 to resolve the halo wake), assumed static potentials (Ibata~\& Razoumov~(1998)
did not allow the halo or the satellite to respond dynamically, artificially
restricting the degrees of freedom), or explored a different 
region of parameter space than that appropriate for the Sgr--Milky Way system (Tsuchiya~(2002) was concerned with the effects of
the LMC, not Sgr).

Our simulations presented here consist of a Milky Way based on model~3 of
Dehnen~\& Binney~(1998). The halo has a mass of 
$8.96 \times 10^{11}\mathrm{~M_{\odot}}$ represented by 1048576 collisionless
particles, while the luminous (disk and bulge) component has a mass of
$4.82 \times 10^{10}\mathrm{~M_{\odot}}$ represented by 1048576 collisionless
particles. The satellite 
is modelled by a single particle with mass
$2 \times 10^9\mathrm{~M_{\odot}}$, at the upper range of the
estimated mass of Sgr (Helmi~\& White~2001). The satellite
was placed at its apogalacticon at 70~kpc and evolved forward using
the GADGET code (Springel, Yoshida,~\& White~2001) on an orbit with a
perigalacticon at 12~kpc (Helmi~\& White~2001) and
inclined 45\deg to the galactic plane. One simulation was performed
with the satellite on a prograde orbit and one with the satellite on a
retrograde orbit (a polar orbit simulation was also run but suffered from
    numerical issues and so is not included).

\begin{figure}
\plotfiddle{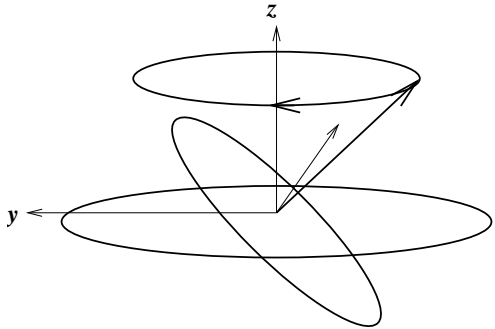}{1.7in}{0}{75}{75}{-100}{0}
\caption{The galactic disk lies in the $xy$ plane. The satellite's angular
  momentum, denoted by the bold vector, initially lies in the $yz$ plane
    and precesses counterclockwise around the $z$ axis.\label{precession}}
\end{figure}
Near perigalacticon, the satellite 
is exposed to the flattened potential of 
the galactic disk, causing its orbit to precess.
In the simulation coordinates, the projection of the satellite's orbital
angular momentum is initially along the $-y$ axis. As it precesses, it
acquires $-L_x$ and $+L_y$ (see Figure~\ref{precession}). In the retrograde
case, all of this angular momentum is acquired by the disk, whereas in
the prograde case the disk only acquires the $x$ component while the
$y$ component is absorbed by the halo. This demonstrates the importance
of including a live halo which can exchange angular momentum with the
other components -- without a live halo, the behaviour of the system
would be qualitatively different.

\begin{figure}
\plottwo{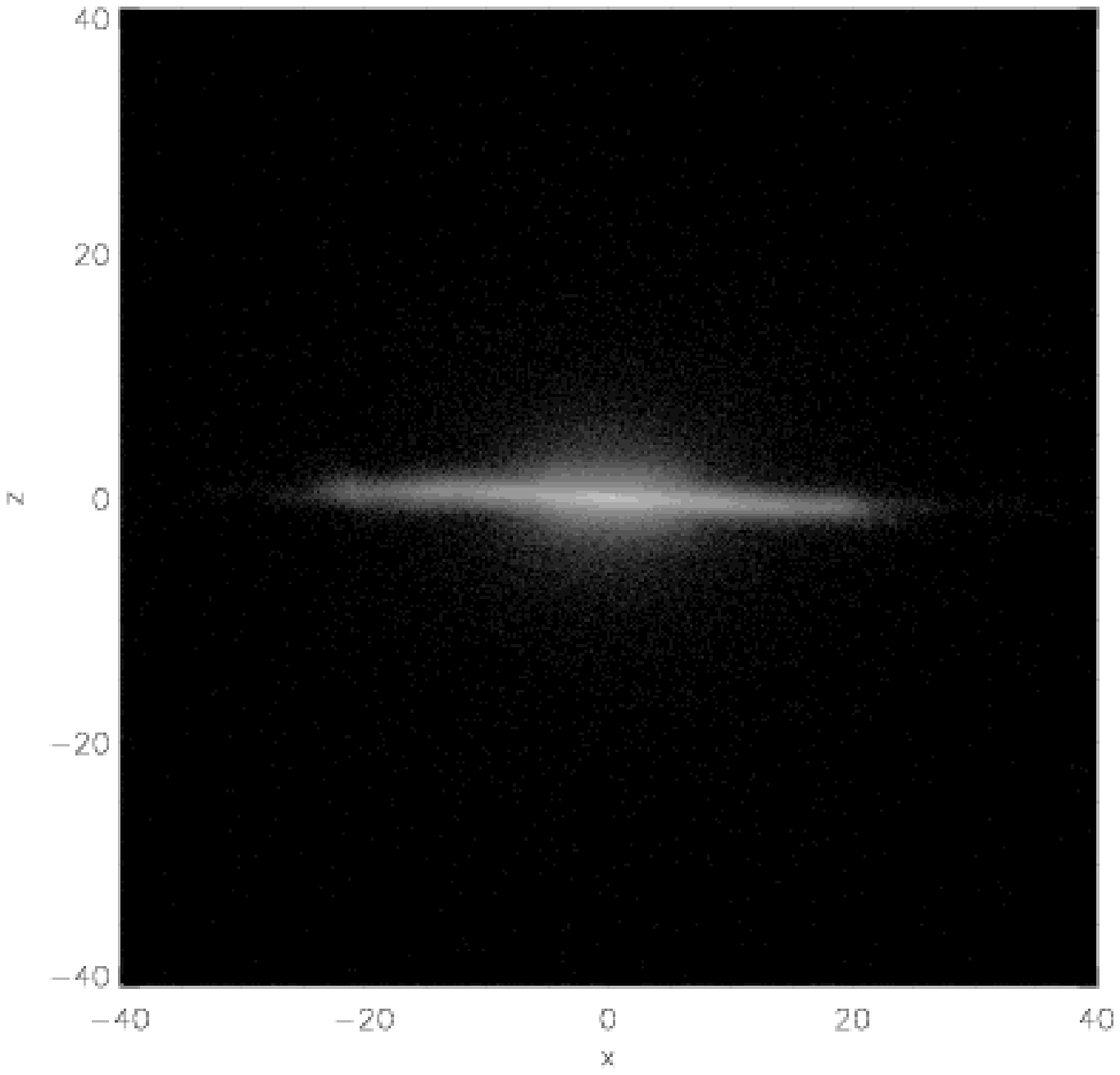}{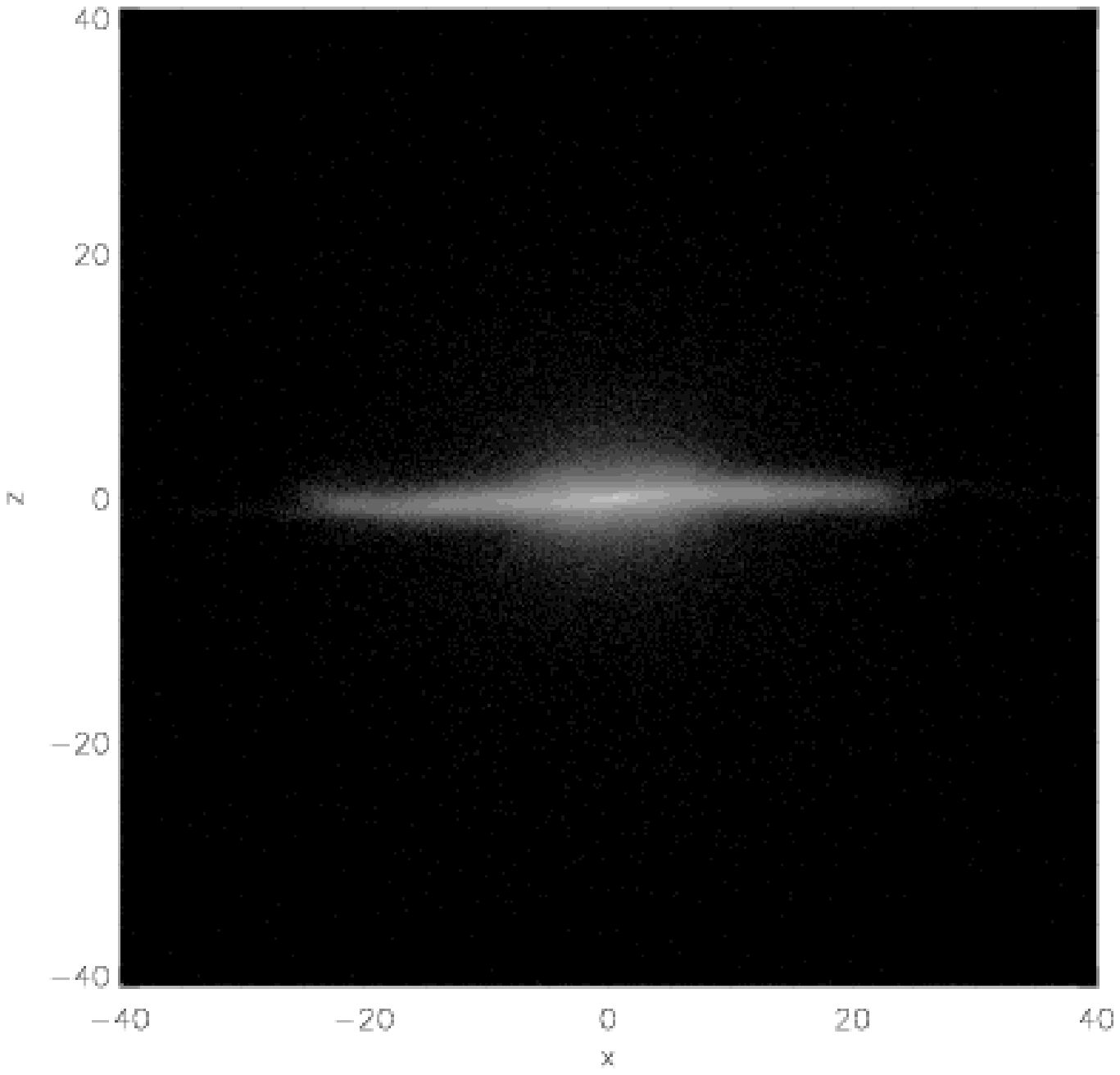}
\caption{Simulation of the Galactic disk after 
 the satellite has orbited for 3.1~Gyr.
  The satellite was placed on a prograde \textit{(left)} (retrograde
  \textit{(right)}) orbit inclined 45\deg to the disk. Intensity
  represents log(projected surface density).\label{disk}}
\end{figure}
Figure~\ref{disk} shows a snapshot of the simulations after 3.1~Gyr, shortly
after the third perigalactic passage. In both cases, 
a warp--like feature has formed. 
The warps are strongest shortly after the perigalactic passages.
The angular momentum transfer is strongly peaked at perigalacticon, causing
the disk to tilt in discrete jerks that send out vertical bending waves
which are seen as warps. These warps are dynamical, and not long-lived.
Evidence that the Milky Way's warp is a dynamical warp comes from
Drimmel~et~al.~(2000), who found that while the Galactic stellar distribution
is warped, the dynamics are not consistent with a stationary or uniformly
precessing warp. However, although the simulated warp is not stationary, 
it appears frequently. 
The warp moves outward at
the group velocity of bending waves in a stellar disk, which
scales as $\pi G \Sigma/\Omega$ for surface density $\Sigma$ and rotational
velocity $\Omega$ (Hofner~\& Sparke~1994),
or $\sim 30\mathrm{~kpc~Gyr^{-1}}$ near the solar circle. Since the period
of Sgr's orbit is approximately 1~Gyr (Helmi~\& White~2001), the warp
reaches the end of the disk shortly before the next satellite passage
warps it anew. Thus if satellites in orbits like Sgr are common around
disk galaxies, a large fraction of disks could appear warped at any given time
even though the warps are not themselves long-lived.


\begin{references}
\reference Bailin, J. 2003, \apj, 583, L79
\reference Dehnen, W., \& Binney, J. 1998, \mnras, 294, 429
\reference Diplas, A., \& Savage, B. D. 1991, \apj, 377, 126
\reference Drimmel, R., Smart, R. L., \& Lattanzi, M. G. 2000, \aap, 354, 67
\reference Freudenreich, H. T., Berriman, G. B., Dwek, E., Hauser, M. G.,
  Kelsall, T., Moseley, S. H., Silverberg, R. F., Sodroski, T. J., Toller,
  G. N., \& Weiland, J. L. 1994, \apj, 429, L69
\reference Garc\`{i}a-Ruiz, I., Kuijken, K., \& Dubinski, J. 2002, \mnras,
  337, 459
\reference Helmi, A., \& White, S. D. M. 2001, \mnras, 323, 529
\reference Hofner, P., \& Sparke, L. S. 1994, \apj, 428, 466
\reference Huang, S., \& Carlberg, R. G. 1997, \apj, 480, 503
\reference Hunter, C., \& Toomre, A. 1969, \apj, 155, 747
\reference Ibata, R. A., \& Razoumov, A. O. 1998, \aap, 336, 130
\reference Ibata, R. A., Wyse, R. F. G., Gilmore, G., Irwin, M. J., \&
  Suntzeff, N. B. 1997, \aj, 113, 634
\reference Lin, D. N. C. 1996, in Gravitational Dynamics, ed. O Lahav,
  E. Terlevich, \& R. J. Terlevich (Cambridge: Cambridge University Press), 15
\reference L\'{o}pez-Corredoira, M., Betancort-Rijo, J., \& Beckman, J. E.
  2002, \aap, 386, 169
\reference Reshetnikov, V. \& Combes, F. 1998, \aap, 337, 9
\reference Shang, Z., Brings, E., Zheng, Z., Chen, J., Burstein, D., Su. H.,
  Byun, Y.-I., Deng, L., Deng, Z., Fan, X., Jiang, Z., Li, Y., Lin, W.,
  Ma, F., Sun, W.-H., Wills, B., Windhorst, R. A., Wu, H., Xia, X., Xu, W.,
  Xue, S., Yan, H., Zhou, X., Zhu, J., Zou, Z. 1998, \apj, 504, L23
\reference Sparke, L. S., \& Casertano, S. 1988, \mnras, 234, 873
\reference Springel, V., Yoshida, N., \& White, S. D. M. 2001, New Astronomy,
  6, 79
\reference Tsuchiya, T. 2002, New Astronomy, 7, 293
\reference Weinberg, M. D., 1998, \mnras, 299, 499
\end{references}
\end{document}